# Anisotropic subwavelength grating perturbation enables zero crosstalk in a leaky mode

Md Faiyaz Kabir[1], Md Borhan Mia[1], Ishtiaque Ahmed[2], Nafiz Jaidye[2], Syed Z. Ahmed[1], and Sangsik Kim[1,2,3,*]

[1]*Department of Electrical and Computer Engineering, Texas Tech University, Lubbock, Texas 79409, USA*

[2]*Department of Physics and Astronomy, Texas Tech University, Lubbock, Texas 79409, USA*

[3]*School of Electrical Engineering, Korea Advanced Institute of Science and Technology, Daejeon 34141, South Korea*

*Corresponding author: sangsik.kim@kaist.ac.kr*

**Electromagnetic coupling via exponentially decaying evanescent field or radiative wave is a primary characteristic of light, allowing optical signal/power transfer but limiting integration density in a photonic circuit. A leaky mode combines both evanescent field and radiative wave, causing stronger coupling and thus not ideal for dense integration. Here we show that a leaky mode with anisotropic perturbation rather can achieve completely zero crosstalk realized by subwavelength grating (SWG) metamaterials. The oscillating fields in the SWGs enable coupling coefficients in each direction to counteract each other, resulting in completely zero crosstalk. We experimentally demonstrate such an extraordinarily low coupling between closely spaced identical leaky SWG waveguides, suppressing the crosstalk by ≈40 dB compared to conventional strip waveguides, corresponding to ≈100 times longer coupling length. This leaky-SWG suppresses the crosstalk of transverse–magnetic (TM) mode, which is challenging due to its low confinement, and marks a novel approach in electromagnetic coupling applicable to other spectral regimes and generic devices.**

Advances in photonic research have led to the integration of various optical components into chip-scale photonic integrated circuits (PICs) for a wide range of applications, including optical computing[1], quantum communication[2,3], light detection and ranging (LIDAR)[4-6], microcomb and optical metrology[7-9], and biochemical sensing[10,11]. The increasing complexity of PICs requires more and more components in a chip, yet chip scalability is limited by the crosstalk prevailing in any optical system. Even a well-confined guided mode (Fig. 1a) exhibits exponentially decaying evanescent fields in the cladding, causing optical coupling between adjacent devices. While this evanescent coupling facilitates some components like couplers and splitters[12-14], it is still the primary origin of crosstalk, limiting the chip integration density. Approaches based on inverse design[15], phase-mismatching[16,17], adiabatic elimination[18], and skin-depth engineering[19,20] have been proposed to reduce the crosstalk, yet they are mostly for transverse electric (TE) mode with additional limiting factors.

Recently, anisotropic metamaterials formed by subwavelength grating (SWG) nano-patterns have been utilized to design various PIC components, greatly expanding the design space[21-23]. By making the grating period smaller than the wavelength, SWGs behave as a homogeneous anisotropic medium, and their effective index can be engineered with their geometric parameters. The large design flexibility of SWGs has been used in advancing various photonic components such as fiber-chip couplers[24-26], optical delay lines[27], biosensors[28,29], Bragg filters[30,31], and polarization controlling devices[32-34]. An extreme skin-depth (eskid) waveguide scheme that utilizes SWGs to reduce the skin depth of evanescent fields was recently proposed, significantly suppressing the crosstalk for dense chip integration[19,20]. However, such a skin-depth suppression approach works only with TE polarization, whose dominant electric field is parallel to the chip surface. In various PICs, transverse-magnetic (TM) mode, whose dominant electric field is vertical to the chip surface, doubles chip capacity and plays important roles in biochemical and gas sensing with its extended fields in the vertical direction[35,36]. Despite its significance, TM is difficult to confine due to a low height-to-width aspect ratio (for easy etching) and exhibits larger crosstalk than TE. The eskid waveguide also causes a stronger coupling for TM mode with increased skin depth[37], and this large TM crosstalk issue still remains a challenge, impeding progress toward high-density chip integration.

As illustrated in Fig. 1b, a leaky mode can be formed by coupling a guided waveguide mode to the continuum of radiation modes in the surrounding infinite clad media[38,39]. While the mode is propagating, the spread of these radiations enables coupling with other devices even when they are far apart. This radiative coupling provides a major advantage in directional couplers[40-42] and polarization splitters[43], as the coupling length remains short with increasing separation distance. But it also proves to counteract when it comes to unwanted waveguide crosstalk, as the cladding radiation significantly enhances the coupling strength between waveguides. Leaky modes, therefore, are not considered ideal for dense integration. However, by orienting the SWGs perpendicular to the propagation direction (Fig. 1c), we can form a leaky mode for TM polarization and achieve zero crosstalk. This counter-intuitive approach relies on the anisotropic nature of SWGs, for which each field component (i.e., $E_x$, $E_y$, and $E_z$) in the radiative waves will be weighted differently than the isotropic cladding case (Fig. 1b). Each component can be engineered anisotropically to cancel out the overall coupling strength by changing the homogenized optical indices of SWG metamaterials. For the practical use of anisotropic leaky mode, the SWG lengths can be truncated finite as in Fig. 1d; still, the anisotropically oscillating fields in the SWG cladding operate the same, and its field perturbations can be engineered depending on the finite width of SWGs that corresponds to the spacing between the two identical waveguides.

In this work, we show that an anisotropic leaky mode realized by SWG metamaterials (as in Fig. 1d) can cancel crosstalk completely, i.e., zero crosstalk. The leaky mode and zero crosstalk are realized with TM polarization, the bottleneck for chip integration due to its lower confinement. Starting by looking into the modal properties of leaky SWG modes, we apply coupled-mode analysis to reveal the unique dielectric perturbation of anisotropic leaky mode, finding zero crosstalk between closely spaced identical SWG waveguides. Then, using Floquet boundary simulations, we design practically implementable SWG waveguides on a standard silicon-on-insulator (SOI) platform and experimentally demonstrate near-zero crosstalk, drastically increasing the coupling length of TM mode by more than two orders of magnitudes.

## Results and Discussion

**Anisotropic leaky mode with SWGs**. To see the modal properties, we first simulated the fundamental TM (TM$_0$) mode of a single waveguide and plotted their field components in each direction. Figures 2a-2c show the cross-sections of the strip, infinite-SWG, and finite-SWG waveguides, respectively. In order to model the anisotropic SWGs, we used the effective medium theory (EMT) with the permittivities $\varepsilon_x = \varepsilon_y = \varepsilon_\parallel$ and $\varepsilon_z = \varepsilon_\perp$ given by[44],

$$\varepsilon_\parallel = \rho \varepsilon_{Si} + (1-\rho)\varepsilon_{air} \tag{1a}$$

$$\varepsilon_\perp = \frac{\varepsilon_{Si}\varepsilon_{air}}{\rho\varepsilon_{air} + (1-\rho)\varepsilon_{Si}} \tag{1b}$$

where $\rho$ is the filling fraction of silicon (Si) in the cladding, and $\varepsilon_{Si}$ and $\varepsilon_{air}$ are the permittivities of Si and air, respectively. The electric field profiles of each waveguide scheme are shown in Figs. 2d-2f, from top to bottom, plotting the normalized Re($E_y$), Re($E_x$), and Im($E_z$) (see Supplementary Information Fig. S1 for magnetic field components). The strip waveguide (Figs. 2a and 2d) supports a well-confined/guided TM$_0$ mode, exhibiting a dominant $E_y$ field. On the other hand, the infinite-SWG waveguide (Figs. 2b and 2e) shows a leaky mode with laterally radiating waves. Now the $E_x$ and $E_z$ are not negligible due to radiative waves, while $E_y$ is still dominant in the core. Truncating the SWG cladding layers to a finite width $w_{swg}$ (Fig. 2c) makes the mode confined, but the oscillating fields in the SWG claddings remain the same exhibiting leaky-like field patterns (Fig. 2f). These oscillating waves in the SWGs can be controlled by changing the $w_{swg}$ (see Supplementary Information Fig. S2), introducing non-trivial dielectric perturbations once coupled with other waveguides. The gap $g$ is introduced between the Si core and SWG claddings to minimize scattering losses from a sharp corner in the experiment, but the modal properties show similar trends even without the gap (see Supplementary Information Fig. S3).

**Zero crosstalk in leaky TM modes.** To examine the coupling effect, we simulated the coupled modes of the two identical waveguides and compared their coupling lengths. The cross-sections and geometric parameters of the coupled strip and SWG waveguides are depicted in Figs. 3a and 3b, respectively. The EMT models in Eq. (1) represent the finite, perpendicular SWG claddings in Fig. 3b. The coupling length $L_c$ is used to quantify the crosstalk, which defines the minimal length over which optical power is maximally transferred from one waveguide to the other[45]. The coupling length is a critical metric for comparing the degree of waveguides crosstalk (i.e., ratio of power exchange), as the degree of crosstalk varies per waveguide length. The simulated effective indices of the coupled $TM_0$ symmetric (orange, $n_s$) and anti-symmetric (blue, $n_a$) modes are plotted in Figs. 3c (strip) and 3d (SWG) as a function of SWG width $w_{swg}$, with their corresponding coupling lengths shown in Figs. 3e and 3f, respectively. The coupling lengths are normalized by the free-space wavelength $\lambda_0 = 1550$ nm, and are evaluated using[45,46],

$$\frac{L_c}{\lambda_0} = \frac{1}{2\Delta n} = \frac{1}{2|n_s - n_a|} \tag{2}$$

where $\Delta n = |n_s - n_a|$ is the index difference between the symmetric and anti-symmetric modes. With the coupled strip waveguides (Fig. 3a), a typical trend where $n_s$ is larger than $n_a$ and they get closer as $w_{swg}$ increases are seen (Fig. 3c), having limited $L_c/\lambda_0$ less than 100 waves (Fig. 3e). This very short coupling length is due to less confinement from $TM_0$ mode for the given separation distances, making $TM_0$ mode difficult for dense integration. For comparison, a typical $L_c/\lambda_0$ of fundamental TE mode for the same separation distance ranges approximately between $10^3$ and $10^4$ waves (see Supplementary Information Fig. S4). However, the coupled SWG waveguides (Fig. 3b) show a non-trivial coupling region where $n_s < n_a$ (gray-shaded region, Fig. 3d). Moreover, at the transition point from the trivial coupling ($n_s > n_a$) to the non-trivial one ($n_s < n_a$), the index difference $\Delta n$ becomes zero ($n_s = n_a$), which indicates infinitely long coupling length $L_c = \infty$ (from Eq. 2). This infinitely long coupling length is directly seen in Fig. 3f. It is worth noting that the $TM_0$ of SWG waveguides supports leaky-like radiative waves in the cladding, which is supposed to exhibit larger crosstalk (thus, less coupling length) unless there is such a non-trivial coupling.

**Anisotropic dielectric perturbation with SWGs.** To further understand the role of SWG leaky mode in achieving zero crosstalk, we investigate each coupling scheme using the coupled-mode analysis[45,46]. The coupling coefficients $\kappa_x$, $\kappa_y$, and $\kappa_z$ from each field component ($E_x$, $E_y$, and $E_z$) are calculated separately and then summed together to get the total coupling coefficient $|\kappa| = |\kappa_x + \kappa_y + \kappa_z|$ (see Methods). Figures 3g and 3h show the calculated coupling coefficients of the coupled strip and SWG waveguides, respectively, as a function of $w_{swg}$: normalized $\kappa_x$, $\kappa_y$, and $\kappa_z$ (dashed lines, left axis) and $|\kappa|$ (solid red line, right axis). The coupling length is calculated using[46],

$$L_c = \frac{\pi}{2|\kappa|} \tag{3}$$

and the corresponding normalized $L_c/\lambda_0$ of the coupled strip and SWG waveguides are plotted in Figs. 3i and 3j, respectively. For a guided $TM_0$ mode (Figs. 3a and 3g), $\kappa_y$ is dominant with a high $E_y$ field, while the other components $\kappa_x$ and $\kappa_z$ are negligible. As $w_{swg}$ enlarges, all the coupling coefficients decrease due to the exponentially decaying evanescent fields in the cladding, reducing the dielectric perturbation strength between the coupled waveguides. On the other hand, in the coupled SWG waveguides (Figs. 3b and 3h), the $\kappa_x$ and $\kappa_z$ show a non-conventional trend, i.e., their magnitudes increase with $w_{swg}$. The oscillating fields in the leaky SWG attributed to this non-conventional dielectric perturbation, which allows the negative $\kappa_x$ and $\kappa_z$ to counteract the positive $\kappa_y$ component, leading to the complete cancellation of the total coupling coefficient $|\kappa| = 0$ at a certain point (Fig. 3h). The corresponding $L_c$ approaches infinity at this $|\kappa| = 0$ point, as seen in Fig. 3j. The results closely match with the full numerical results in Fig. 3f. The shaded regions in Figs. 3h and 3j show the non-trivial coupling regimes where $\kappa < 0$, which corresponds to the $n_s < n_a$ region in Figs. 3d and 3f. Note that this exceptional coupling achieving zero crosstalk is due to the anisotropic dielectric perturbations of the leaky mode realized by SWGs, where $\Delta\varepsilon_x = \Delta\varepsilon_y > \Delta\varepsilon_z$. With a conventional isotropic ($\Delta\varepsilon_x = \Delta\varepsilon_y = \Delta\varepsilon_z$) leaky mode, such a complete zero crosstalk is impossible to achieve as $|\kappa|$ is always greater than zero. These coupling singularities via anisotropic dielectric perturbations could also vary with different core widths $w$, as shown in Supplementary Information Fig. S5.

**Experimental results**. In order to verify our findings, we fabricated the coupled SWG waveguides and experimentally characterized their crosstalks. We fabricated our SWG devices on a 220 nm-thick SOI wafer using a standard electron beam nanolithography process (see Methods). Figure 4a shows the scanning electron microscope (SEM) images of the fabricated devices with a schematic experimental setup for measuring the crosstalk. As the ideal EMT model and practical SWGs would differ in effective indices, we used Floquet modal simulations to optimize structures with realistic parameters (see Methods). Figure 4b shows schematics of the simulation domains (top: perspective-view and bottom: top-view), and Fig. 4c represents the mode profiles ($E_y$) of the coupled $TM_0$ symmetric (top) and anti-symmetric (bottom) modes. Figure 4d shows the simulated crosstalk spectra of the coupled SWG waveguides (solid lines) for different core widths $w$ = 565 nm (red), 570 nm (blue), and 575nm (green). For comparison, the crosstalks of the coupled strip waveguides without SWGs are also plotted (dashed lines). As expected from previous modal simulations using an ideal EMT, complete zero crosstalks (dips) are seen, resulting in infinitely long coupling lengths as in Fig. 4e. For the experimental characterization, we sent light $I_0$ through one of the coupled waveguides and measured output power ratios $I_2/I_1$, which defines the crosstalk. Grating couplers are used for interfacing the chip and fibers. Figures 4f and 4g show the experimentally characterized crosstalk and $L_c/\lambda_0$ corresponding to the results in Figs. 4d and 4e, respectively. The crosstalk of coupled SWG waveguides is drastically suppressed down to as low as -50 dB (Fig. 4f), approximately 40 dB lower than coupled strip waveguides. In terms of the coupling length (Fig. 4g), the maximum $L_c/\lambda_0$ of the SWG waveguides is ≈ $10^4$ waves, which is about two orders of magnitudes longer than the strip case. Unlike the ideal SWG simulations, there is a practical limit in measuring the minimum crosstalk due to background noise in the chip, either from sidewall roughness scattering or cross-coupling at the strip to SWG transition. Still, to our knowledge, the $TM_0$ crosstalk suppression shown here is the lowest recorded, with a coupling length encompassing ≈$10^4$ waves. To explicitly show the effectiveness of our approach, we summarize key performance factors in Table 1 and compare them with other TM crosstalk suppression approaches[15,18,47-49].

# Conclusion

In summary, we uncovered that a leaky mode can achieve complete zero crosstalk by anisotropically engineering dielectric perturbations to cancel out the couplings from each field component. We realized such an anisotropic leaky mode using the perpendicularly arrayed SWGs and optimized via Floquet numerical simulations. We experimentally demonstrated the extreme suppression of TM crosstalk on an SOI platform, achieving ≈40 dB crosstalk suppression and two orders of magnitude longer coupling lengths than typical strip waveguides. Our work directly provides a practical and easily applicable waveguide platform for overcoming the integration density limit of TM mode and should be pivotal for advancing PIC technologies in applications like on-chip biochemical/gas sensing and polarization-encoded quantum/signal processing. Furthermore, our proposed method of using anisotropic SWGs to achieve zero crosstalk reveals a novel coupling mechanism with a leaky mode, easily extendable to other integrated photonics platforms and covering visible to mid-infrared and terahertz wavelengths beyond the telecommunication band.

## Methods

**Coupled-mode analysis.** The coupling coefficient components $\kappa_i$ of the coupled strip and SWG waveguides were calculated using the coupled-mode theory[45,46],

$$\kappa_i = \frac{\omega \varepsilon_0}{4} \iint \Delta\varepsilon_i(x,y) E_{1i}(x,y) E_{2i}^*(x,y) dx dy \qquad (4)$$

where $i = x$, $y$, and $z$ denotes the coupling coefficient from each field component. By isolating the waveguides at each side (without coupling), the unperturbed normalized electric fields of the $TM_0$ modes are obtained as $E_{1i}$ and $E_{2i}$. $\Delta\varepsilon_i$ is the dielectric perturbation imposed by the presence of the individual waveguides on each other. The total coupling coefficient $|\kappa|$ between the coupled waveguides was obtained by adding the individual components together,

$$|\kappa| = |\kappa_x + \kappa_y + \kappa_z| \qquad (5)$$

and the corresponding coupling length is given by $L_c = \pi/(2|\kappa|)$. The analysis was carried out at a free-space wavelength of $\lambda_0 = 1550$ nm.

**Floquet modal simulations.** To model and simulate the practically implementable SWG waveguides, alternating Si layers are arranged perpendicular to the waveguide propagation direction (z-axis). The structure is spatially repeated with period $\Lambda = 100$ nm by imposing Floquet boundary conditions at each end of the simulation domain (see Fig. 4b). For setting the Floquet periodicity, we defined the wave vector as $k_z = \frac{2\pi}{\lambda} n_{\text{eff}}$, where $n_{\text{eff}}$ is the effective index of the $TM_0$ mode at a particular wavelength $\lambda$. The simulations were carried out for different core widths indicated by $w = 565$ nm (red), 570 nm (blue), and 575 nm (green) in Figs. 4d and 4e. The Floquet simulations can reasonably estimate the geometric parameters required to achieve complete zero crosstalk. These optimized parameters are fixed at height $h = 220$ nm, SWG width $w_{\text{swg}} = 570$ nm, and gap $g = 65$ nm with a filling fraction of 0.45. Edges of SWGs are also rounded, considering the fabricated devices shown in the SEM images.

**Device fabrication.** The photonic chips were fabricated on an SOI wafer with 220 nm thick Si and 2 μm $SiO_2$ substrate, using the JEOL JBX 6300-fs electron beam lithography (EBL) system. The operating conditions were 100 KeV energy, 400 pA beam current, and 500 μm x 500 μm field exposure. A solvent rinse was done initially, followed by $O_2$ plasma treatment for 5 min. Hydrogen silsesquioxane resists (HSQ, Dow-Corning XR-1541-006) was spin-coated at 4000 rpm and pre-exposure baked on a 90° hotplate for 5 min. The exposure dose used was 2800 μC/cm². During shot shape writing, the machine grid shape placements, the beam stepping grid, and the spacing between dwell points were 1 nm, 4 nm, and 4 nm, respectively. The resist was developed in 25% tetramethylammonium hydroxide (TMAH) heated to 80° and placed into the solution for 30 s, and then rinsed in flowing deionized water for 2 min and isopropanol for 10 s. Nitrogen was blown in for air drying. The die was placed in an $O_2$ plasma asher at 100 W for 15 s with 10 sccm $O_2$ flowing into the system. The unexposed top Si device layer was etched using Trion Minilock III ICP-RIE etcher at 50 W RF power and 6.2 mTorr pressure with $Cl_2$ and $O_2$ gas flowing into the chamber at 50 sccm and 1.4 sccm, respectively. An active cooling system maintained the stage temperature stably at 10°C during the entire etching process.

**Crosstalk characterization.** The crosstalk of the strip and SWG coupled waveguides was characterized by measuring their respective output power ratio. Light from a tunable laser source with optical power $I_0$ was coupled to the input port using grating couplers (see Fig. 4a). A Keysight Tunable Laser 81608A was used as a source, and an angle-polished (8°) fiber array was used to couple light into the grating coupler. A polarization controller was used to ensure the input light polarization was TM. By simultaneously measuring the output powers $I_1$ and $I_2$ at the through and coupled ports, the crosstalk was calculated as the ratio $I_2/I_1$. A Keysight N7744A optical power meter was used to detect the output powers. The coupling length $L_c$ was extracted from the relation[45],

$$\frac{I_2}{I_1} = \tan^2\left(\frac{\pi L}{2L_c}\right), \tag{6}$$

where $L = 30$ μm is the length of the coupled waveguides. The measurements were taken for core widths $w$ = 580 nm (red), 585 nm (blue), and 590 nm (green) (see Figs. 4f and 4g).

## Acknowledgments

This material is based upon work supported by the National Science Foundation under Grants No. 2144568 and No. 1930784. This work was performed, in part, at the Center for Integrated Nanotechnologies (CINT), an Office of Science User Facility operated for the U.S. Department of Energy Office of Science by Los Alamos National Laboratory and Sandia National Laboratories. S.K. acknowledge the support from the KAIST new faculty research fund and KSEA young investigator grant.

## Author Contributions

S.K. developed the idea and supervised the project. M.K. and M.M. modeled the structures and performed the numerical simulations with the support of S.K. I.A. fabricated the Si devices. N.J. and S.Z.A. experimentally characterized the devices. S.K. and M.K. prepared the figures and wrote the manuscript.

## Competing interests

The authors declare no competing interests.

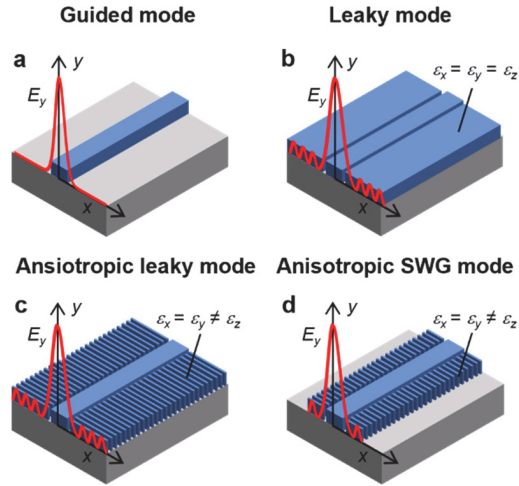

**Fig. 1. Waveguide configuration with subwavelength gratings (SWGs).** Waveguide schematics under evaluation (blue: Si and grey: SiO$_2$). The red lines illustrate the fundamental TM$_0$ modes ($E_y$). **a,** A typical strip waveguide supporting a guided mode with exponentially decaying evanescent fields. **b,** Placing infinite slabs adjacent to the strip results in leaky mode with a radiative loss into the slabs. **c,** A perpendicular array of infinite SWGs can replace the slab, supporting a leaky mode, but SWGs provide anisotropic field oscillations. **d,** By truncating the SWGs, the mode will be guided without radiative losses while preserving its leaky-like oscillations in the anisotropic SWG claddings. When coupled with other waveguides, this leaky-like anisotropic oscillation exhibits a non-conventional anisotropic perturbation and can result in zero crosstalk.

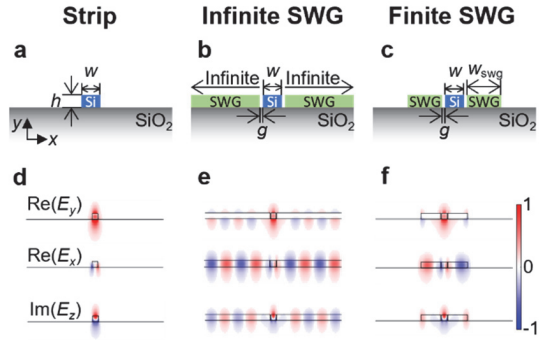

**Fig. 2. Modal properties of the strip, infinite subwavelength grating (infinite-SWG), and finite-SWG waveguides. a-c,** Cross-sections of the strip, infinite-SWG, and finite-SWG waveguides, respectively. The SWGs are represented by the equivalent model using the effective medium theory. **d-f,** Mode profiles of $TM_0$ modes in each waveguide scheme: (**d**) strip, (**e**) infinite-SWG, and (**f**) finite-SWG. From top to bottom, $Re[E_y]$, $Re[E_x]$, and $Im[E_z]$ are plotted. The mode profiles exhibit (**d**) guided mode, (**e**) leaky mode, and (**f**) a hybrid mode with oscillating patterns. The geometric parameters are $h$ = 220 nm, $w$ = 600 nm, $g$ = 65 nm, and $w_{swg}$ = 2 µm.

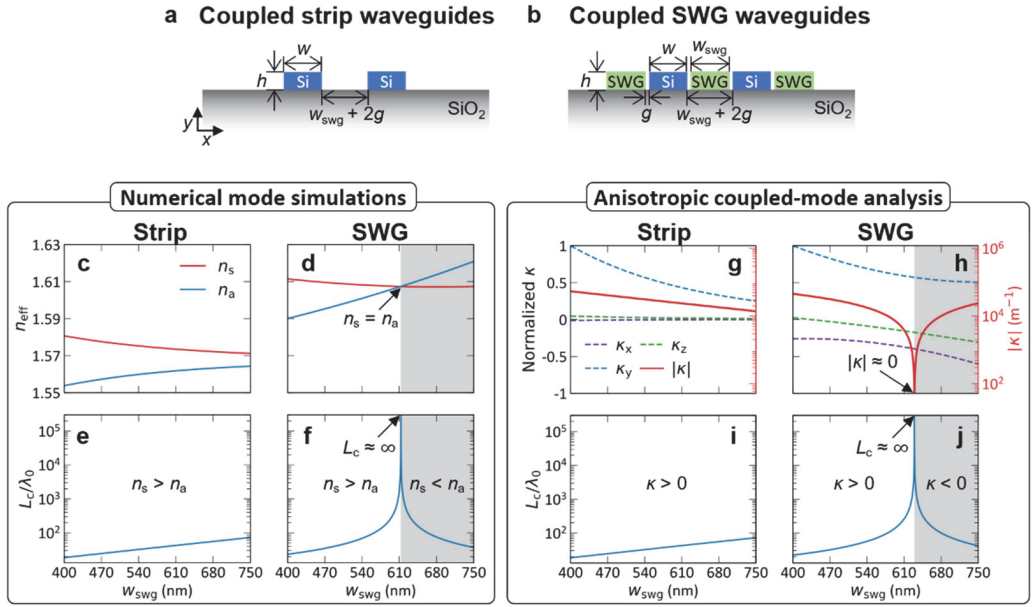

**Fig. 3. Zero crosstalk in TM$_0$ mode with coupled SWG waveguides. a, b,** Schematics of the coupled (**a**) strip and (**b**) SWG waveguides. **c, d,** Numerically simulated effective indices of the coupled symmetric ($n_s$, red) and anti-symmetric ($n_a$, blue) TM$_0$ modes for (**c**) strip and (**d**) SWG waveguides, and (**e, f**) their corresponding normalized coupling lengths $L_c/\lambda_0$. **g, h,** Normalized coupling coefficients $\kappa_x$ (purple dashed), $\kappa_y$ (blue dashed), $\kappa_z$ (green dashed), and the total coupling coefficient $|\kappa|=|\kappa_x+\kappa_y+\kappa_z|$ (red solid). **i, j,** Corresponding $L_c/\lambda_0$ for the coupled strip and SWG waveguides, respectively. The grey-shaded areas represent the non-trivial coupling region, where (**d, f**) $n_a > n_s$ and (**h, j**) $\kappa < 0$. The free-space wavelength is $\lambda_0$ = 1550 nm, and the other parameters are $h$ = 220 nm, $w$ = 530 nm, and $g$ = 65 nm.

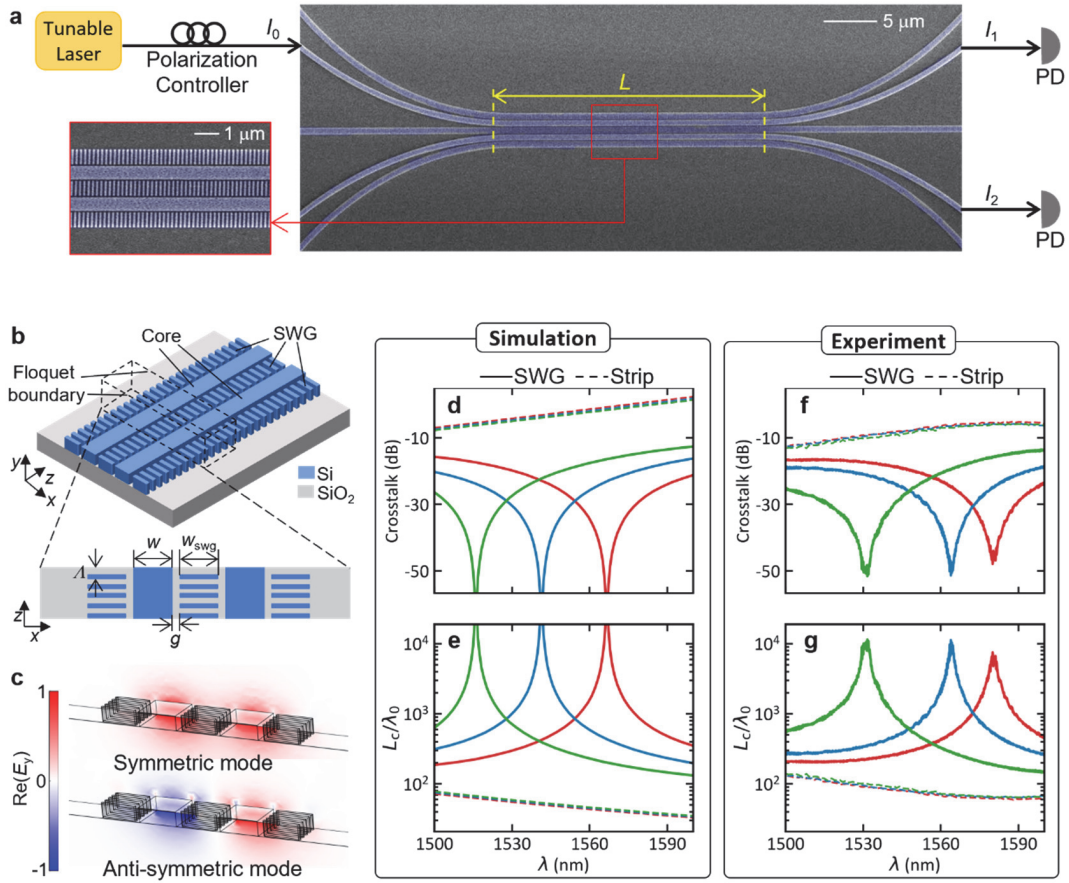

**Fig. 4. Experimental demonstration of near-zero crosstalk in $TM_0$ mode with coupled SWG waveguides**. **a,** SEM images of the fabricated devices with experimental setup for measuring the crosstalk. The zoomed-in image shows the coupled SWG waveguides. **b,** 3D schematic and top view of the simulation domain. The structure is periodically repeated in the propagation direction (z-axis) with period $\Lambda$ = 100 nm using the Floquet boundary condition. **c,** Simulated mode profiles of the symmetric and anti-symmetric modes. **d, e,** Numerically simulated (**d**) crosstalk and (**e**) corresponding normalized coupling length $L_c/\lambda_0$ of the strip (dashed) and SWG (solid) waveguides: $w$ = 565 nm (red), 570 nm (blue), and 575nm (green). Other parameters are fixed to $h$ = 220 nm, $w_{swg}$ = 570 nm, $L$ = 30 µm, and $g$ = 65 nm with a filling fraction of 0.45. **f, g,** Experimentally characterized (**f**) crosstalk and (**g**) corresponding $L_c/\lambda_0$ of the strip (dashed) and SWG (solid) waveguides: $w$ = 580 nm (red), 585 nm (blue), and 590 nm (green).

**Table 1. Coupling length $L_c$ comparison of TM mode with different waveguide configurations.**

| Waveguide Configuration | Polarization | H x W (nm x nm) | Separation (μm) | Coupling Length, $L_c$ (mm) | Optical Loss |
|---|---|---|---|---|---|
| Strip[47-49] | TM | 220 x 590 | 0.80 – 1.20 | 0.010 – 2.75 | Low (0.40 – 0.74 dB/cm) |
|  | TE | 220 x 450 |  |  |  |
| Cloaking[15] | TM | 300 x 300 | 0.80 | 0.72 | High (>300 dB/cm) |
| Adiabatic Elimination[18] | TM | 340 x 220 | 0.94 | 5.90* | N/A |
| **Leaky-SWG (This Work)** | TM | **220 x 590** | **1.19** | **15.50** | Low (≈3 dB/cm)[19,27] |

*The wavelength of ref.[18] is at 1310 nm, while all the other wavelengths are at 1550 nm.

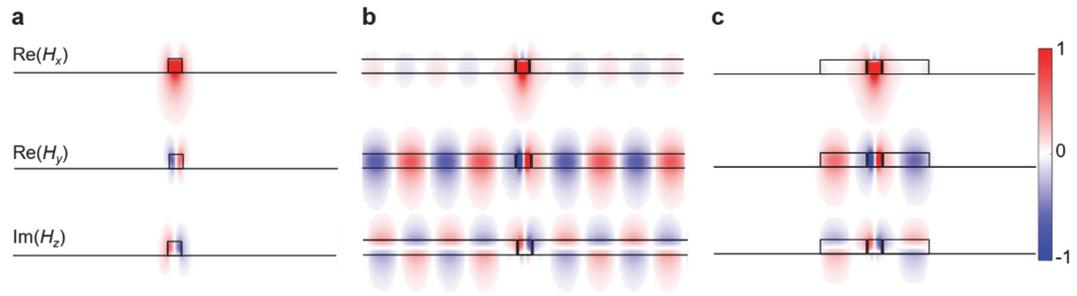

**Fig. S1. Magnetic field profiles of the strip, infinite subwavelength grating (infinite-SWG), and finite-SWG waveguides. a-c,** Magnetic field profiles of the fundamental TM modes ($TM_0$) in each waveguide scheme: **a,** strip, **b,** infinite-SWG, and **c,** finite-SWG. From top to bottom, Re[$H_x$], Re[$H_y$], and Im[$H_z$] are plotted. The geometric parameters are the same as in Fig. 2 of the main manuscript.

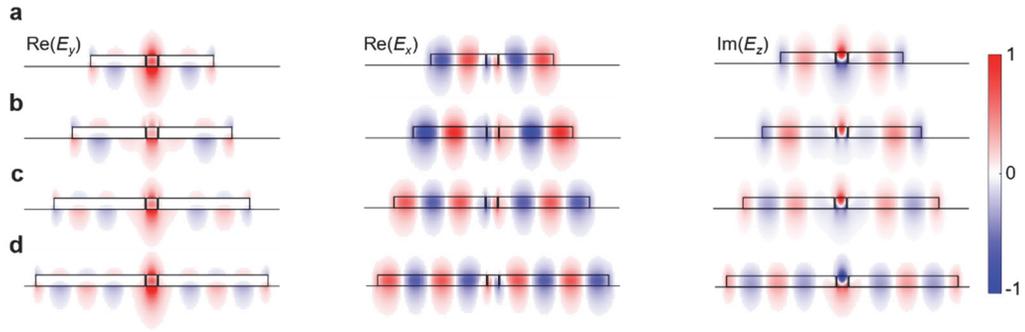

**Fig. S2. Mode profiles with $w_{swg}$ variation for finite-SWG waveguides. a-d,** Mode profiles of fundamental TM modes ($TM_0$) for different SWG widths: **a,** $w_{swg}$ = 3 µm, **b,** $w_{swg}$ = 4 µm, **c,** $w_{swg}$ = 5 µm, and **d,** $w_{swg}$ = 6 µm. From left to right, Re[$E_y$], Re[$E_x$], and Im[$E_z$] are plotted. The oscillations of the anisotropic leaky mode can be controlled by changing the SWG width, as seen from the different radiation patterns. The other geometric parameters are the same as in Fig. 2 of the main manuscript.

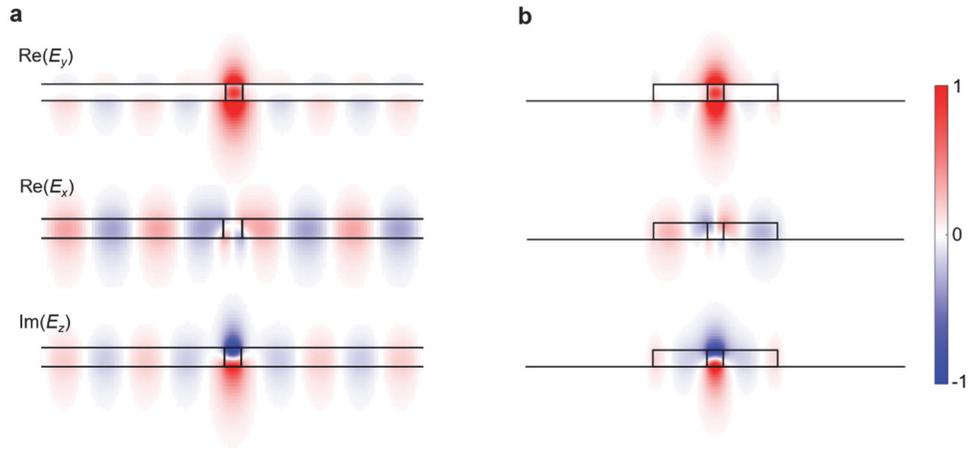

**Fig. S3. Modal properties of the infinite subwavelength grating (infinite-SWG) and finite-SWG waveguides without the gap between the core and the cladding. a, b,** Mode profiles of fundamental TM modes (TM$_0$) in (**a**) infinite-SWG and (**b**) finite-SWG. From top to bottom, Re[$E_y$], Re[$E_x$], and Im[$E_z$] are plotted. The mode profiles exhibit (**a**) a leaky mode for the infinite-SWG and (**b**) a hybrid mode with an oscillation for the finite-SWG, even without the gap between core and SWG cladding. The geometric parameters are the same as in Fig. 2.

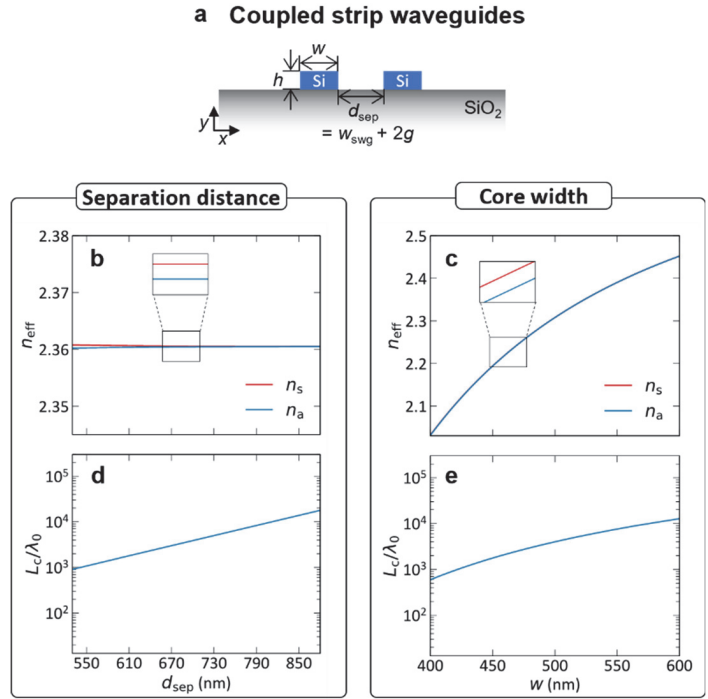

**Fig. S4. Modal simulation results of TE$_0$ mode for coupled strip waveguides. a,** Schematic of the coupled strip waveguides. **b, c,** Numerically simulated effective indices of the coupled symmetric ($n_s$, red) and anti-symmetric ($n_a$, blue) TE$_0$ modes for strip waveguides with variation in **b,** separation distance $d_{sep}$, and **c,** core width $w$, and, **d, e,** their corresponding normalized coupling lengths $L_c/\lambda_0$. Due to the large confinement of the TE$_0$ modes, the coupling length is significantly higher than the TM$_0$ modes (Fig. 3e) for the same separation distance $d_{sep} = w_{swg} + 2g$.

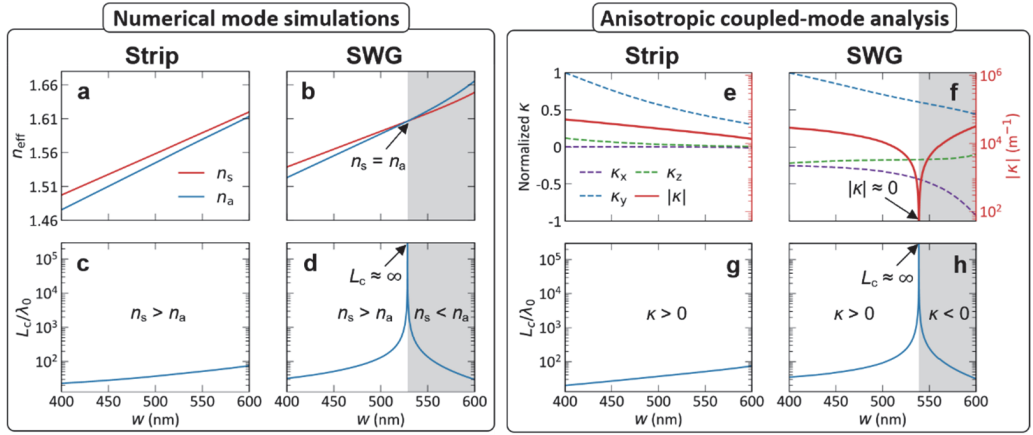

**Fig. S5. Coupling length vs. core width *w* variation in $TM_0$ mode with coupled SWG waveguides. a, b,** Numerically simulated effective indices of the coupled symmetric ($n_s$, red) and anti-symmetric ($n_a$, blue) $TM_0$ modes for **a**, strip and **b**, SWG waveguides, and **c, d,** their corresponding normalized coupling lengths $L_c/\lambda_0$. All the parameters and simulations are similar to Fig. 3 but as a function of the core width *w*. **e, f,** Normalized coupling coefficients $\kappa_x$ (purple dashed), $\kappa_y$ (blue dashed), $\kappa_z$ (green dashed), and the total coupling coefficient $|\kappa|=|\kappa_x+\kappa_y+\kappa_z|$ (red solid). **g, h,** Corresponding $L_c/\lambda_0$ for the coupled strip and SWG waveguides, respectively. The grey-shaded areas represent the non-trivial coupling region, where (**b, d**) $n_a > n_s$ and (**f, h**) $\kappa < 0$. The free-space wavelength is $\lambda_0$ = 1550 nm, and the other parameters are *h* = 220 nm, $w_{swg}$ = 620 nm, and *g* = 65 nm.